\documentclass[conference]{IEEEtran}
\IEEEoverridecommandlockouts
\usepackage{cite}

\usepackage{amsmath,amssymb,amsfonts}
\usepackage{graphicx}
\usepackage{subcaption}
\usepackage{enumitem}
\usepackage{booktabs}

\usepackage{algorithm}
\usepackage{algpseudocode}
\usepackage{textcomp}
\usepackage{xcolor}
\def\BibTeX{{\rm B\kern-.05em{\sc i\kern-.025em b}\kern-.08em
    T\kern-.1667em\lower.7ex\hbox{E}\kern-.125emX}}
\begin{document}

\title{Metering traffic flows for perimeter control through auction-based signalling using connected vehicles\\

\thanks{The authors would like to acknowledge the financial contribution of the EU Horizon Europe Research and Innovation Programme, Grant Agreement No. 101103808 ACUMEN.}
}

\author{\IEEEauthorblockN{1\textsuperscript{st} Alexander Roocroft}
\IEEEauthorblockA{\textit{Department of Transport and Planning} \\
\textit{Faculty of Civil Engineering and Geosciences}\\
\textit{Delft University of Technology}\\
Delft, The Netherlands \\
a.roocroft@tudelft.nl, ORCID: 0000-0002-6551-1800}
\and
\IEEEauthorblockN{2\textsuperscript{nd} Marco Rinaldi}
\IEEEauthorblockA{\textit{Department of Transport and Planning} \\
\textit{Faculty of Civil Engineering and Geosciences}\\
\textit{Delft University of Technology}\\
Delft, The Netherlands \\
m.rinaldi@tudelft.nl}
}

\maketitle

\begin{abstract}
Urban traffic congestion remains a critical challenge in modern cities, with traffic signal control systems often struggling to manage congestion during peak travel times. Perimeter control of a Protected Network (PN) has emerged as a potential solution to reducing gridlock in urban networks. This paper proposes a novel auction-based mechanism for green time allocation at signalized intersections, for effective perimeter control application. Utilising a Sealed Bid, Second Price auction framework, our approach combines real-time traffic monitoring with market-inspired mechanisms to regulate vehicle inflows into PN areas. Unlike existing methods that focus primarily on gated links, our system allocates budgets to individual traffic movements, providing greater flexibility in managing multi-directional flows. We evaluate the proposed mechanism using a test case intersection with a single controlled inflow, comparing it against a volume-based fixed-time approach. The results demonstrate that our auction-based method controls flows into the PN with improved accuracy, outperforming the volume-based approach in terms of inflow regulation, queue management and delays. The framework can be applied in real time to any generic intersection, offering a scalable solution for urban traffic management. This work bridges the gap between perimeter control and market-based intersection auctions, providing a pathway for further research on adaptive traffic management systems.
\end{abstract}


\section{Introduction}

Urban traffic congestion is a critical challenge in modern cities, significantly impacting mobility, safety, and environmental quality. During peak travel times, conventional traffic management systems often fail to prevent the oversaturation of key urban areas, leading to gridlock. Research by \cite{GAYAH2014} has demonstrated that locally adaptive signal control schemes have limited effectiveness during heavy congestion and gridlock, primarily due to downstream bottlenecks and queue spillbacks. As an alternative, perimeter control has emerged as a promising solution. By leveraging real-time traffic monitoring and dynamic adjustments, perimeter control can manage vehicle inflows into protected networks (PNs), drawing on foundational work on the Network Fundamental Diagram (NFD) in \cite{DAGANZO2007} and \cite{GEROLIMINIS2008}. These studies have shown that maintaining optimal traffic density through perimeter control can help avoid gridlock and improve overall network efficiency.

Much of the existing perimeter control research focuses on determining the permissible flows into a PN during specific time intervals to maximize throughput. Techniques such as reinforcement learning (\cite{YU2025}), Proportional-Integral (PI) controllers (\cite{KEYVANEKBATANI2021}), and game theory (\cite{Elouni2021}) have been explored to calculate these flows. However, a critical yet understudied aspect is the integration of flow metering into the traffic signal control systems that form the gated perimeter. Many studies simplify this by adding an extra traffic light after the intersection to meter gated flows (\cite{KEYVANEKBATANI201374}), but this approach is impractical due to queue spillbacks that block the preceding intersection. 

An alternative method to this, which is also used, combines metering with locally adaptive traffic signals by adjusting green times for gated links \cite{KEYVANEKBATANI2021}. It converts the metered flow volumes to green times using cycle length and road capacity. Although relatively easy to implement, it has significant limitations, focusing on single gated links and lacking flexibility for multi-directional flow control. Furthermore, most research on this approach has focused on its broader impact on traffic metrics within the PN, while its ability to precisely regulate controlled flows remains under-explored.

Within intersection control, a promising alternative approach that has gained attention is the use of market-based auctions. A key recent work \cite{Iliopoulou2022}, has applied advanced auction-based mechanisms with important properties, including incentive compatibility, configuration flexibility, and prevention of starvation (i.e., ensuring all users eventually receive green time). It assumes the widespread use of connected vehicles with reliable Vehicle-to-Infrastructure (V2I) communication and sufficient penetration to control intersections, potentially provided by smartphone technology. By incorporating heterogeneous user data, such as urgency levels through value-of-time (VOT), into a Sealed Bid, Second Price (SBSP) auction framework, the approach has shown significant enhancements in key traffic performance metrics, including reduced delays and improved flow rates, when compared to conventional fixed-time control. However, based on a review of the literature, it appears that such auction frameworks have not yet been applied to perimeter control.

In this paper, we bridge this gap by integrating perimeter control into state-of-the-art market-based intersection auctions. We propose a novel method for green time allocation that combines dynamic user bids with flow limits in a SBSP auction mechanism. Our approach builds on that employed in \cite{Iliopoulou2022} to include flow control while maintaining the desirable properties of market-inspired mechanisms (e.g. flexibility, fairness). Unlike existing methods that focus on traffic volumes at gated links, our auction-based approach can allocate flow budgets to individual traffic directions and movements. This enables greater flexibility in managing multi-directional flows and supports a wider range of intersection applications. 

We evaluate the proposed mechanism using a test case intersection, comparing it against the volume-based fixed-time approach of \cite{KeyvanEkbatani23122019}. The results demonstrate that our method provides improved queue management and flow regulation, enabling more precise metering for perimeter control. 

The remainder of the paper is structured as follows. Firstly, the methodology of the proposed mechanism is described. Then, the proposed mechanism is compared with volume-based fixed-time control using simulation. Lastly, a final section concludes the work and provides directions for future research.

\section{Methodology}

\subsection{Auction framework}

In our framework, traffic signal phases (i.e., groups of signals controlling different movements) compete in an auction to determine the next phase to receive green time. Each phase is represented by a wallet agent that collects bids from road users and submits them to an intersection manager. The manager then conducts a computationally efficient auction and implements the winning phase through signal control.

In this section, each intersection consists of a set of lanes, \( L \), that accommodate \( M \) movements, controlled by a set of signal phases, \( S \). Each phase \( s \in S \) serves a subset of movements, \( M_s \subset M \), with the corresponding subset of lanes serving that phase denoted as \( L_s \subset L \). For clarity, more of the notation used in our framework is defined in Table \ref{tab:notation}.

\begin{table}[h!]
\centering
\begin{tabular}{cl}
Symbol & Definition \\
\hline
$A^{t}$ & Set of bidding phases at time $t$ \\
$C^{t}$ & Set of bids at time $t$ \\
$D_{s}^{t}$ & Set of bidding distances for phase $s$ at time $t$ \\
$K_{l}^{t}$ & Set of vehicles in lane $l$ at time $t$ \\
$L$ & Set of lanes $l$ \\
$\bar{L}^t$ & Subset of all active lanes $l$ at time $t$ \\
$L_{s}^t$ & Subset of active lanes serving phase $s$ at time $t$ \\
$M$ & Set of movements $m$ at intersection \\
$M_{s}$ & Subset of all movements serving phase $s$ \\
$\hat{M}_{s,\gamma}$ & Subset of all movements using inflow $\gamma$ in $M_{s}$ \\
$\bar{M}_s^t$ & Subset of movements active for phase $s$ at time $t$ \\

$N_{l}$ & Set of bidding vehicles per lane $l$ \\
$S$ & Set of phases $s$ \\
$\hat{s}$ & Currently active phase \\
$T$ & Set of time steps $t$ in time horizon \\
$\Gamma$ & Set of inflow roads $\gamma$ \\
$b_{i}^{t}$ & Bid amount for bidding vehicle $i$ at time $t$ \\
$d_{l}^{t}$ & Bidding distance per lane $l$ at time $t$ \\
$e_{s}^{t}$ & Active duration of phase $s$ at time $t$ \\
$G_{s}^{\text{max}}$ & Maximum green duration for phase $s$ (60 s) \\
$G_{s}$ & Minimum green duration for phase $s$ (3 s)\\
$G^{ext}$ & Green time extension for repeat winning phase (3 s)\\
$m_{i}$ & Movement of bidding vehicle $i$ \\
$s_{i}^{t}$ & Traffic light phase for bidding vehicle $i$ at time $t$ \\
$VOT_{i}$ & Value of time for user of bidding vehicle $i$ \\
$Y_{s}$ & Yellow phase duration corresponding to green phase $s$ (2 s)\\
$P_{i}(w_{i}^t)$ & Impatience function for user of bidding vehicle $i$ \\
$q_{l}^{t}$ & Number of vehicles in queue in lane $l$ at time $t$ \\
$r_{i'}^{t}$ & Auction payment of user $i'$ in winning phase at time $t$\\
$\rho$ & Saturation headway (2 s)\\
$v_{i}^{t}$ & Speed of bidding vehicle $i$ at time $t$ \\
$w_{i}^{t}$ & Waiting time for bidding vehicle $i$ at time $t$ \\
$x_{i}^{t}$ & Distance from intersection of bidding vehicle $i$ at time $t$ \\
$\lambda_{min}$ & Minimum vehicle length \\
$\lambda_{max}$ & Maximum vehicle length \\
$t_p^{start}$ & Start time of perimeter activation \\
$t_p^{end}$ & End time of perimeter activation \\
$f_{\gamma,z}^{max}$ & Maximum allowable flow for inflow $\gamma$ in budget period $z$ \\
$f_{\gamma}^t$ & Flow count for inflow $\gamma$ at time $t$ \\

\end{tabular}
\caption{Notation and Definitions. Default simulation values are indicated in brackets.}
\label{tab:notation}
\end{table}

\subsubsection{Bidding distance determination}
The vehicles that take part in the auction are determined by setting a bidding distance from the intersection, with any vehicle within the limit allowed to participate. At each time step \textit{t}, a new auction commences and a bidding distance factor $z$ is determined based on the average waiting time for each lane $l$ with a red light indication. Subsequently, the proposed bidding distance per lane is calculated based on this bidding distance factor. The largest of the lane bidding distances for a specific phase \textit{s} is applied to all vehicles waiting in active lanes for that phase. Minimum and maximum permissible distances are applied to make sure that at least one vehicle can bid and bidding vehicles can physically cross the intersection.  The process is described in Algorithm 1. 

\begin{algorithm}
\caption{Bid Distance Calculation}\label{alg:1}
\begin{algorithmic}[1] 
    \State Determine active lanes $\bar{L}_s^t$ serving movements $\bar{M}_s^t$ $\forall s$.
    \For{each phase \( s \in A^t \)}
        \State \textbf{if} \( e_s^t = 0 \) \textbf{then}
        \For{each lane \( l \) in \( \bar{L}_s^t \)}
            \State \textbf{if} \( q_l^t > 0 \) \textbf{then}
                \State \( z_l^t = \frac{\sum_{i \in K_l^t} w_i^t}{q_l^t} , \quad l \in \bar{L}_s^t \)
            \State \textbf{else}
                \State \( z_l^t = 0 , \quad l \in \bar{L}_s^t \)
            \State \textbf{end if}
        \EndFor
    \EndFor

    \For{each phase \( s \in A^t \)}
        \State \textbf{if} \( e_s^t > 0 \) \textbf{then}
        \For{each lane \( l \) in \( \bar{L}_s^t \)}
            \State \( d_l^t = \hat{d}^{b}, \quad l \in \bar{L}_s^t \)
            \State \( d_l^t \to D_s^t \)
        \EndFor
        \State \textbf{else}
        \For{each lane \( l \) in \( \bar{L}_s^t \)}
            \State \( d_{min}^s = |\bar{L}_s^t| * \lambda_{min} \)
            \State \( d_{max}^s = (G_s/\rho) * \lambda_{max} \)
            \State \( d_l^t = d_{min}^s + \left(d_{max}^s-d_{min}^s\right) \frac{z_l^t}{\sum_{\lambda\in \bar{L}^t}z_\lambda^t}, \quad l \in \bar{L}_s^t \)
            \State \( d_l^t \to D_s^t \)
        \EndFor
        \State \textbf{end if}
        \For{each lane \( l \) in \( \bar{L}_s^t \)}
            \State \( d_l^t = \frac{\arg\max(D_s^t)}{|\bar{L}_s^t|}, \quad l \in L \)
        \EndFor
    \EndFor
    
\end{algorithmic}
\end{algorithm}

\subsubsection{User bid calculation}
In our SBSP auction, users are assumed to have a known VOT and adjust their bids dynamically based on their accumulated waiting time at the intersection. As waiting time increases, users become more impatient and raise their bids according to an impatience function. The winning phase $W$ is selected based on the total bids ${c}^t_s$ for each phase $s$. The runner-up $Z$ is then selected as the phase that has the highest bid below the winning phase's bid. The payments $r_{i'}^t$ of the users $i'$ in the winning phase are then scaled with the ratio of the runner-up to the winning phases' total bids. 

The auction mechanism is adapted from the approaches described in \cite{Iliopoulou2022} and \cite{Roocroft2025}, for further details on the bid calculation process, refer to these works.

\subsection{Flow control}

When perimeter control is active ($t_p^{start}\leq t \leq t_p^{end}$), vehicle entry into the PN is restricted. The duration of this active period can be adjusted to mitigate congestion risks during peak demand times when gridlock is most likely. An external perimeter control algorithm, such as a PI controller \cite{KEYVANEKBATANI2021}, regulates vehicle entry into the PN through designated gated intersections. It dynamically adjusts entry rates over budget periods of length $\tau$ to minimize congestion and maximize throughput.

\begin{figure}[t]

    \centering
    \includegraphics[width=.3\textwidth]{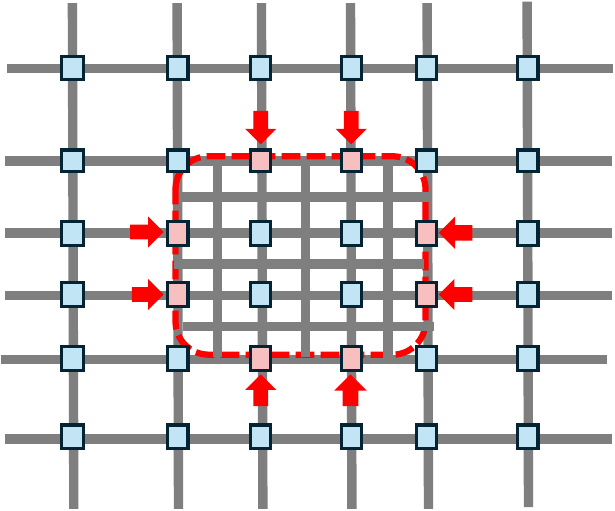} 

    \caption{Illustration of a road network with a protected network region (adapted from \cite{Elouni2021}). The grid layout of roads (grey lines) has 36 intersections indicated by the boxes. The blue boxes are intersections without flow control, the red boxes are intersections implementing flow control into the protected network (PN) (area within red dashed line). The red arrows indicate the roads that contain inflows into the PN. }
    \label{fig:PerimeterControlExample}

\end{figure}

We define a set of inflows, $\Gamma$, representing road sections where traffic enters the gated intersection before continuing to the PN. Figure \ref{fig:PerimeterControlExample} shows an illustration of a grid network with a PN and associated inflows (red arrows). In our case, the method is applied to inflows from one direction only, but could also be adapted to restrict vehicles from multiple inflows (e.g., the corner intersections of the PN). 

Using the external limit on PN traffic entry, a budget is assigned for the traffic movements that use the inflow road to enter the PN. We define, for each inflow road $\gamma \in \Gamma$, a maximum allowable vehicle count $f_{\gamma,z}^{max}$ for a budget period $z$ of length $\tau$. These parameters can be varied depending on the operation of the external perimeter control algorithm. For example, a longer length of budget period and a higher maximum allowable vehicle count would allow more vehicles to enter the PN.

Each inflow $\gamma$ has an associated set of movements $\hat{M}_{s,\gamma}$ from different phases $s$. Initially, the auctions proceed with all movements participating, such that the set of active movements at time $t$ is $\bar{M}_s^t = M_s$, where $M_s$ is the set of all movements serving phase $s$. At time, $t$, the value of $f_{\gamma}^t$ tracks how many vehicles have passed through the inflow $\gamma$ in the current budget period $z$ (e.g., measured by a loop detector). If $f_{\gamma}^t$ exceeds $f_{\gamma,z}^{max}$, then all $m \in \hat{M}_{s,\gamma}$ for all phases $s$ are excluded from the active movements $\bar{M}_{s}^t$ participating in the subsequent auctions of that budget period. They remain excluded until the end of the budget period, when $f_{\gamma}^t$ resets to zero and the active movements of each phase $\bar{M}_s^t$ are restored to $M_s$.

\begin{figure}[t]
    \centering
    \begin{subfigure}[b]{0.35\textwidth} 
        \centering
        \includegraphics[width=\textwidth]{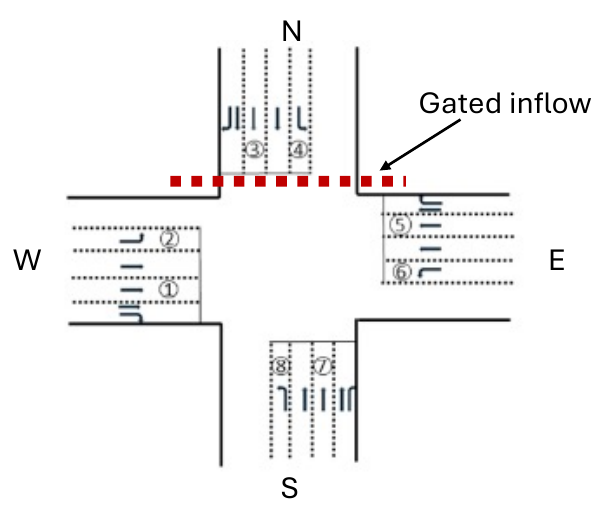} 
        \caption{Intersection Layout}
        \label{fig:1a}
    \end{subfigure}

    \vspace{1em} 

    \begin{subfigure}[b]{0.3\textwidth} 
        \centering
        \includegraphics[width=\textwidth]{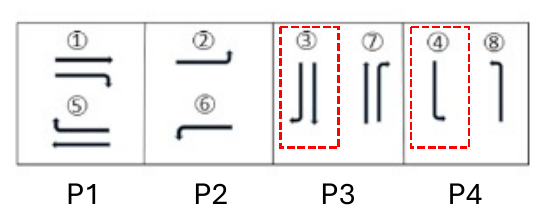} 
        \caption{Phase Diagram of Movements}
        \label{fig:sub2}
    \end{subfigure}

    \caption{Layout of the test case intersection with phase diagram (adapted from \cite{Iliopoulou2022}). The North arm is the gated inflow when the perimeter control is active. This corresponds to movement 3 in phase P3 and movement 4 in phase P4, highlighted in red.}
    \label{fig:testcase}
\end{figure}

An example intersection in Figure \ref{fig:testcase} illustrates the process. The intersection has four arms, with the North arm, $N$, serving as the gated inflow, such that $\Gamma = \{N\}$. Figure \ref{fig:testcase}(b) presents the phase diagram of all movements at the intersection, \( M_s \). The movements associated with the North inflow, movement 3 in phase P3 and movement 4 in phase P4 (highlighted in red), form the sets \( \hat{M}_{P3,N} = \{3\} \) and \( \hat{M}_{P4,N} = \{4\} \). If, during the budget period $z$, the count on the North inflow exceeds its maximum threshold ($f_{N}^t > f_{N,z}^{max}$), then the corresponding movements are removed from the set of active movements for the remaining auctions for that period. Consequently, the active movements in phases P3 and P4 are updated to $\bar{M}_{P3}^t = \{7\}$ and $\bar{M}_{P4}^t = \{8\}$, while movements from other phases continue to participate. After the end of the budget period $z$, all the movements participate once again in the next auction at the start of the following budget period. 

Our auction mechanism including flow control is detailed in Algorithm 2. When the perimeter control is inactive, the auction proceeds without the flow control.

\begin{algorithm}
\caption{Traffic Signal Auction with Flow Control}\label{alg:2}
\begin{algorithmic}[1]
    \State $t = 0$, $z = 0$, $\hat{t}=\tau$
    \State Set $G_s$, $G_s^{max}$, $G^{ext}$, $Y_s$, $A^0 = S$, $f_{\gamma,0}^{max}$, $\bar{M}_s^0 = M_s \forall s$
    \While{$t < T$}
        \State Determine bidding vehs. per lane $N_l$ based on Alg. 1
        \State Run second-price sealed-bid auction
        \State \hspace{1em} Collect bids from bidding vehs. using Eqs. (1), (2)
        \State \hspace{1em} Calculate bids for each phase,
        \Statex \hspace{3.5em} $c_s^t = \sum_{i \mid x_i^t \le d_l^t, i \in N_l, m_i \in \bar{M}_s} b_i^t, \quad c_s^t \in C^t, s \in A^t$
        \State \hspace{1em} Determine auction winner from bids,
        \Statex \hspace{3.5em} $W = \arg\max (C^t), \quad W \in A^t$
        \State \hspace{1em} Determine the runner-up to the winning phase,
        \Statex \hspace{3.5em} $Z = \arg\max (C^t \backslash W), \quad Z \in A^t$
        \State \hspace{1em} Determine payments of users $i'$ in phase $W$, 
        \Statex \hspace{3.5em} 
        $r_{i'}^t = b_{i'}^t \cdot \frac{c_Z^t}{c_W^t}$
        \If{W $\neq$ $\hat{s}$}
            \State Activate yellow light for movements $M_{\hat{s}}$ for $Y_{\hat{s}}$ 
            \State $t = t + Y_{\hat{s}}$
            \State Activate green light for movements $M_W$ for $G_W$ 
            \State $t = t + G_W$
            \State $\hat{s} = W$
            \State $A^t = S$
        \ElsIf {W $=$ $\hat{s}$}
            \State Extend green light for movements $M_W$ for $G^{ext}$
            \If{$e_{W}^t > G^{max}_{W} : A^t = S\backslash W$}
                \State $t = t + G^{ext}$
            \EndIf
        \EndIf
        \Statex \hspace{1.25em} Activation of perimeter control, 
        \If{$t_p^{start}\leq t \leq t_p^{end}$}
        \For{each inflow \( \gamma \in \Gamma \)}
             \If{$f_{\gamma }^t > f_{\gamma ,z}^{max}$}
                \State Exclude movements exceeding budget,
                \Statex \hspace{7em} $\bar{M}_s^t = M_s\backslash \hat{M}_{s,\gamma}\quad ,  \forall s$
             \EndIf
              \State $\hat{t} = \hat{t} - 1$
             \If{$\hat{t} = 0$}
                 \Statex \hspace{4.2em} Reset for next budget period,
                 \State $z=z+1$
                \State $\hat{t}=\tau$, $f_{\gamma } = 0$, $\bar{M}_s^t = M_s \quad \forall \gamma ,s$
            \EndIf
        \EndFor
        \EndIf
    \EndWhile
    \State Terminate
\end{algorithmic}
\end{algorithm}

\section{Application}
\subsection{Test Case}

The four-way intersection presented in Figure \ref{fig:testcase}, with its layout and signal phases, is used as a test case for the performance of our framework. Each arm consists of four incoming and four outgoing lanes spanning 750 m. There are four phases, each with two movements that can be independently signalled. For modelling the perimeter control, we consider the North arm to be the gated inflow. When active during the second hour of the simulation ($t_p^{start} =1$, $t_p^{end} = 2$), the budget period $\tau$ is 5 mins.

All simulations run for three hours using the default values from Table \ref{tab:notation}. The VOT follows a uniform distribution between 20 and 40 euros/ hr. The impatience parameters, \( \alpha_{1} \) and \( \alpha_{2} \), are uniformly distributed within \([0.1, 0.5]\) and \([20, 60]\), respectively.

To assess the impact of maximum allowable flow count on performance, we analyse multiple limits assuming equal lane flows. The flow on each arm is set at 800 vehicles per hour modelled using a Poisson distribution, with a turning percentage of 15\% for left and right lanes. Each lane has a capacity of 900 vehicles per hour. We test the following hourly limits on the inflow: 100 veh/hr, 250 veh/hr, 400 veh/hr, 550 veh/hr, 700 veh/hr. In a budget period $\tau$ of 5 mins, these correspond to a maximum allowable vehicle count $f_{\gamma ,z}^{max}$ of 8, 20, 33, 45, 58 vehicles, respectively.

As a benchmark, the performance is compared with the volume-based, fixed-time adjustment approach described in \cite{KeyvanEkbatani23122019}. It uses a volume-based calculation to convert the metered flows from the North inflow into green times within a fixed-time cycle when the perimeter control is active.

We evaluate the different methods using AIMSUN, a microscopic traffic simulator widely used for modelling control strategies \cite{Aimsun2024}. Each simulation is run for 10 replications, from which the calculated average is used for the results.

\begin{figure}[t]
    \centering
    \begin{subfigure}{0.35\textwidth}
        \centering
        \includegraphics[width=\textwidth]{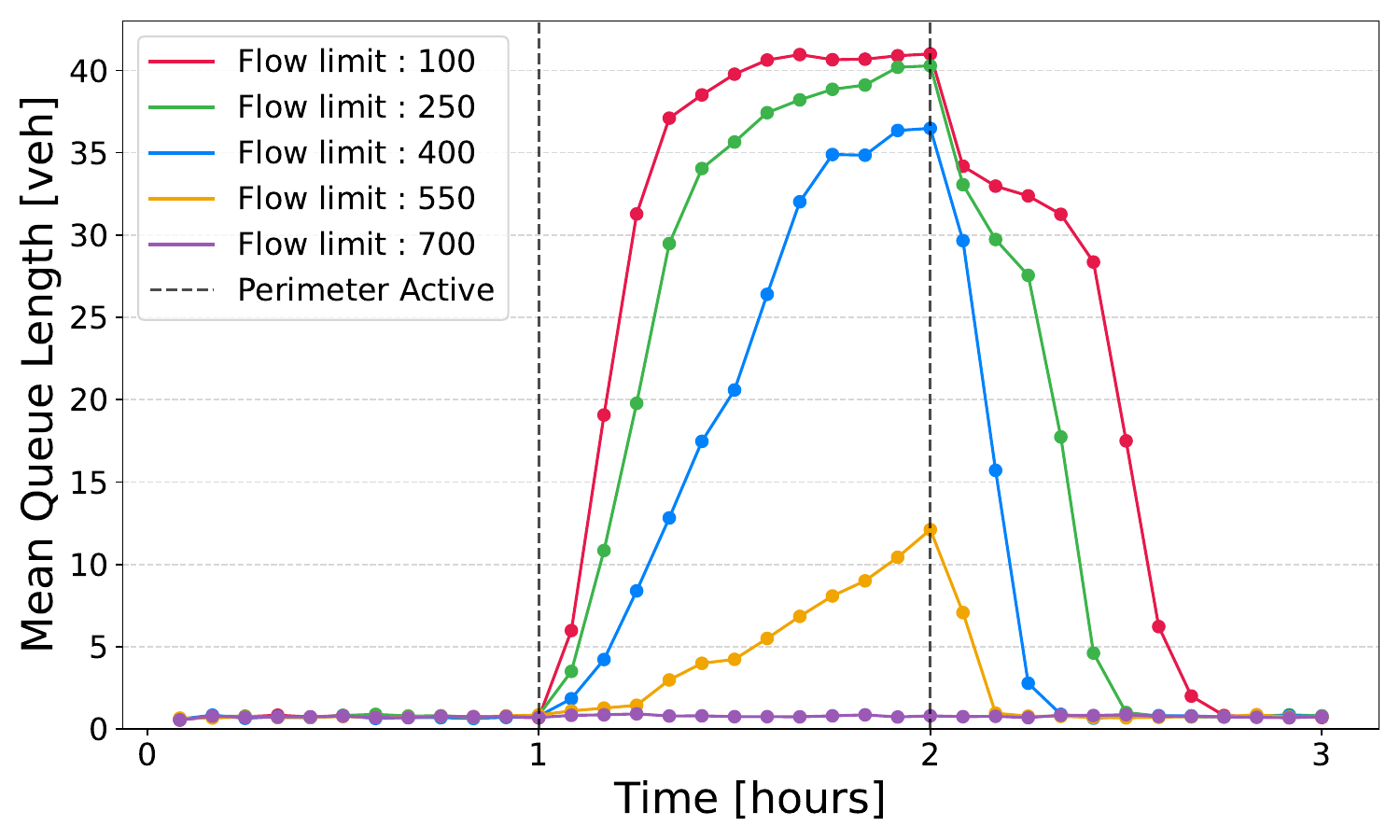} 
        \caption{Auction-based}
        \label{fig:MeanQueue_a}
    \end{subfigure}
    \hfill
    \begin{subfigure}{0.35\textwidth}
        \centering
        \includegraphics[width=\textwidth]{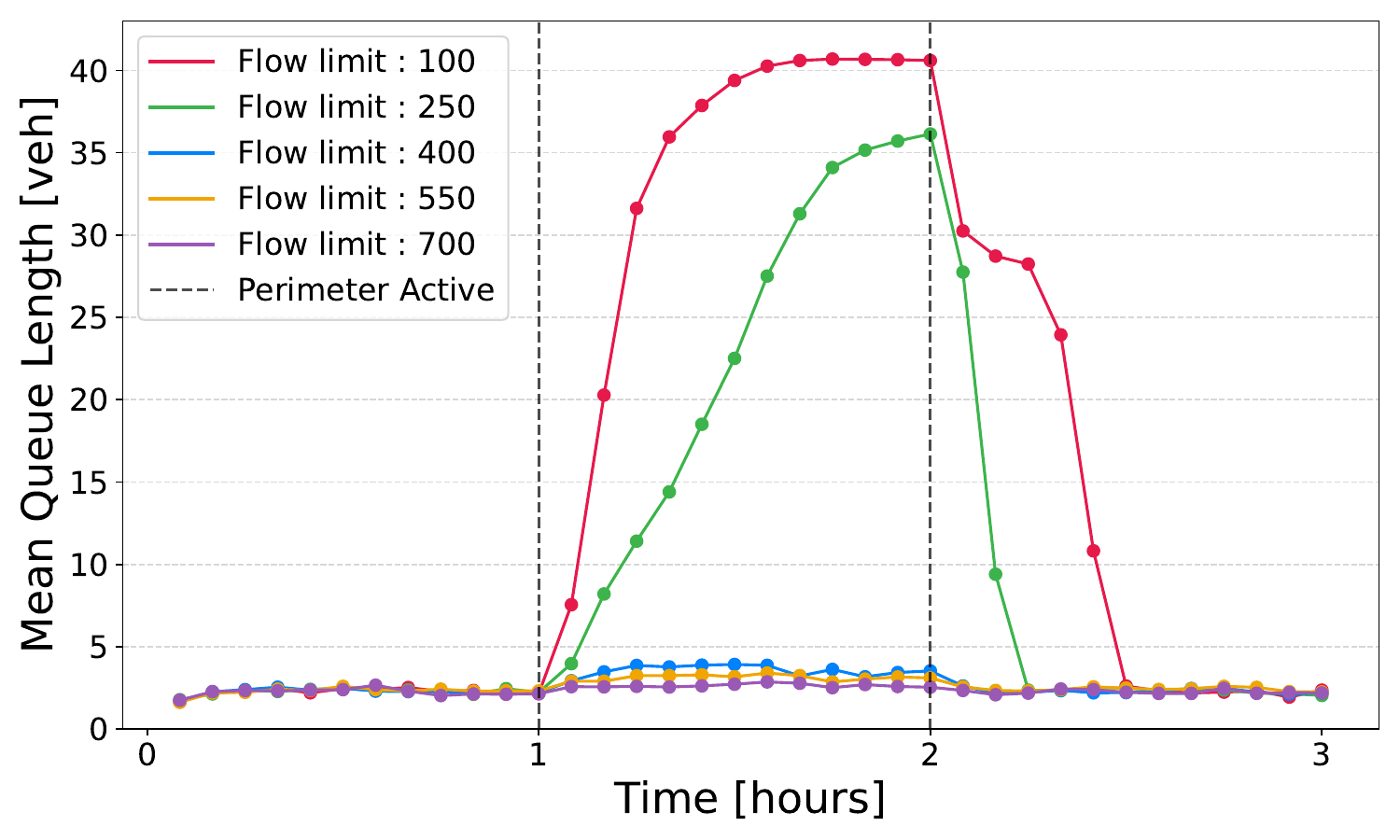} 
        \caption{Volume-based}
        \label{fig:MeanQueue_b}
    \end{subfigure}
    \caption{Mean queues on North arm for every five minutes of simulation. Perimeter control active from hours 1 to 2.}
    \label{fig:MeanQueue}
\end{figure}

\begin{figure}[t]
    \centering
    \begin{subfigure}{0.35\textwidth}
        \centering
        \includegraphics[width=\textwidth]{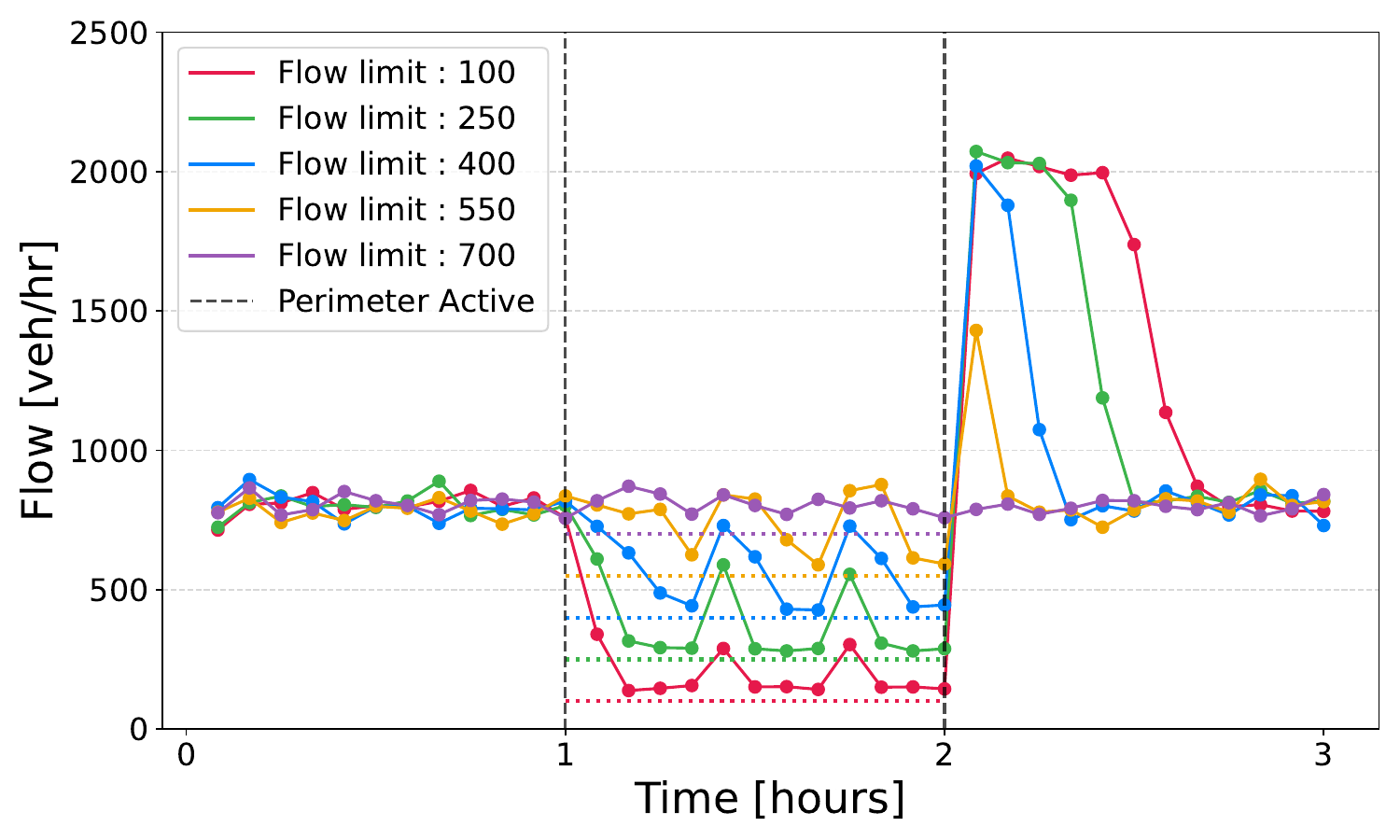} 
        \caption{Auction-based}
        \label{fig:Inflow_a}
    \end{subfigure}
    \hfill
    \begin{subfigure}{0.35\textwidth}
        \centering
        \includegraphics[width=\textwidth]{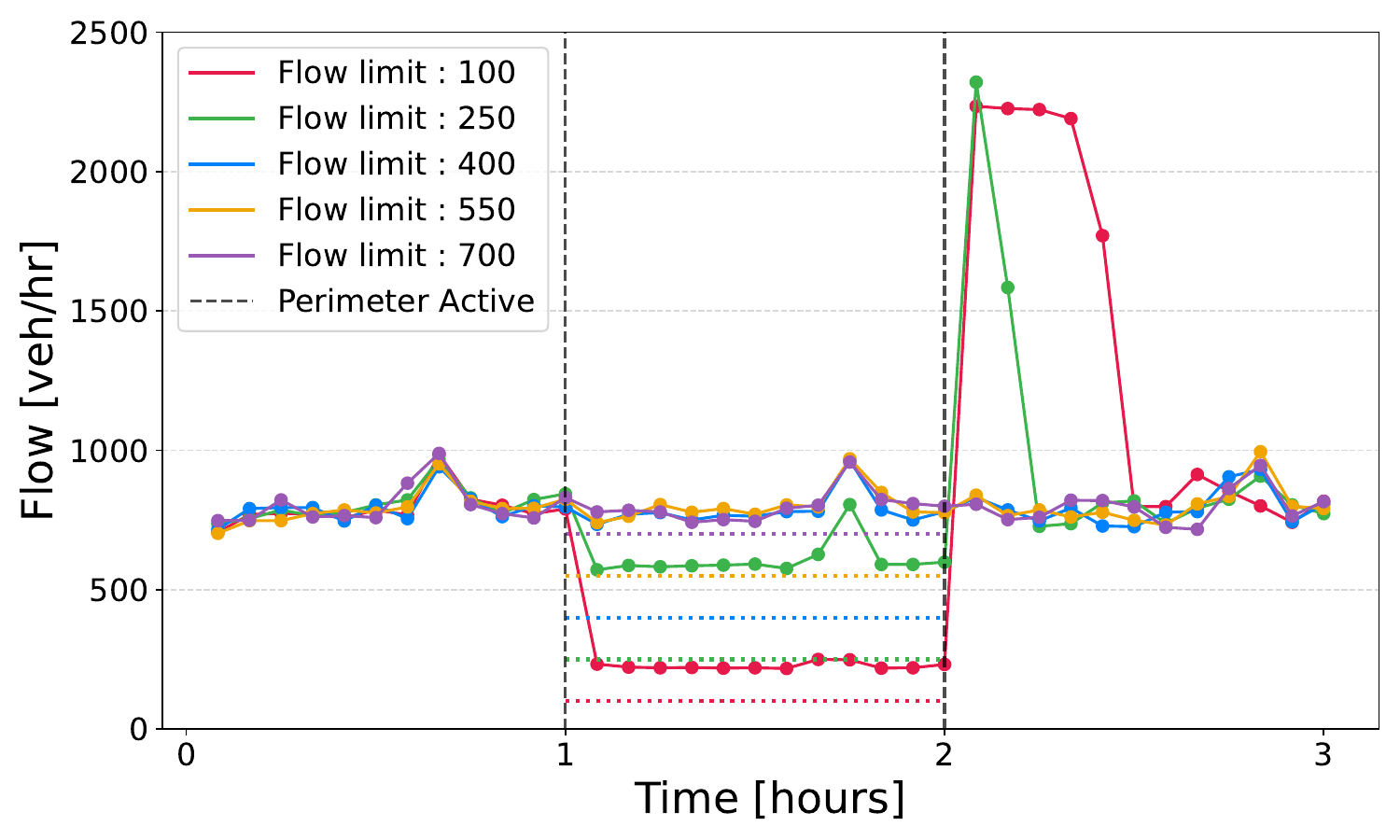} 
        \caption{Volume-based}
        \label{fig:Inflow_b}
    \end{subfigure}
    \caption{Mean inflow from North arm for every five minutes of simulation. Perimeter control active from hours 1 to 2. Dotted lines indicate the target flow limits, solid lines indicate the delivered inflows.}
    \label{fig:Inflow}
\end{figure}

\subsection{Results}

To evaluate the suitability for perimeter management of our auction-based mechanism against the volume-based benchmark, we analyse queue formation and inflow regulation during perimeter activation. We also assess the difference in impact on delays for traffic passing through the intersection.

\subsubsection{Queue Formation}
Figure \ref{fig:MeanQueue} presents the mean queue lengths in 5-minute intervals for both approaches. During perimeter activation, the auction-based approach (Figure \ref{fig:MeanQueue_a}) demonstrated effective queue management, with queue lengths increasing with reduced flow limits. This indicates successful traffic restriction from the North arm. For most flow limits, queue formation followed the expected pattern of queues growing steadily during the control period, approaching the road’s maximum storage capacity (approximately 40 vehicles), and dissipating fairly quickly after the control was lifted, returning to baseline levels by hour 3. However, under the weakest flow limit (700 veh/hr), the approach struggled to restrict inflows, resulting in queue patterns similar to unrestricted conditions. This suggests that the auction-based method is less effective at higher (i.e., less restrictive) flow limits.

In contrast, the volume-based approach (Figure \ref{fig:MeanQueue_b}) showed limited effectiveness, with significant queue formation only occurring under the most restrictive flow limits (100 and 250 veh/hr). For higher limits, queues remained relatively unchanged, indicating poor flow regulation. These results highlight the auction-based method’s superior ability to provide queue management across a wider range of flow limits.
\subsubsection{Inflow Regulation}
The difference between the two approaches for North arm inflows during perimeter activation can be seen in Figure \ref{fig:Inflow}. Consistent with effective queue management, the auction-based approach (Figure \ref{fig:Inflow_a}) successfully regulated inflows for all tested limits except the highest (700 veh/hr), often closely adhering to the target values indicated by the dotted lines. In contrast, the volume-based approach (Figure \ref{fig:Inflow_b}) struggled to enforce flow limits above 250 veh/hr, with inflows mostly exceeding the targets, consistent with poor queue management. 

Table \ref{tab:volume_auction_flow_limits} provides comparison statistics for inflows, showing that the auction-based method achieved, on average, closer alignment with the target limits than volume-based. The difference is most pronounced when examining the minimum inflows. For most of the perimeter activation period, the auction-based inflows remained close to the minimum values, typically within 30 to 40 veh/hr of the target limit. This significantly outperformed the volume-based approach, which consistently failed to meet the target limits. The remaining difference between the minimum values of the auction-based approach and target limits is likely due to minimum green time constraints, which restrict the precision of the control.

\begin{table*}[t]
    \centering
    \resizebox{\textwidth}{!}{%
    \begin{tabular}{lccc ccc ccc ccc ccc}
        \toprule
        & \multicolumn{3}{c}{Flow Limit = 100} 
        & \multicolumn{3}{c}{Flow Limit = 250} 
        & \multicolumn{3}{c}{Flow Limit = 400} 
        & \multicolumn{3}{c}{Flow Limit = 550} 
        & \multicolumn{3}{c}{Flow Limit = 700} \\
        \cmidrule(lr){2-4} \cmidrule(lr){5-7} \cmidrule(lr){8-10} \cmidrule(lr){11-13} \cmidrule(lr){14-16}
        & Min & Max & Mean & Min & Max & Mean & Min & Max & Mean & Min & Max & Mean & Min & Max & Mean \\
        \midrule
        Auction & 138.0 & 340.0 & 188.5 & 280.0 & 610.0 & 365.4 & 427.0 & 730.0 & 559.8 & 589.0 & 877.0 & 738.3 & 758.0 & 871.0 & 808.3 \\
        Volume & 217.2 & 249.6 & 226.5 & 571.2 & 805.2 & 607.7 & 734.4 & 963.6 & 783.5 & 740.4 & 969.6 & 801.8 & 741.6 & 957.6 & 797.0 \\
        \bottomrule
    \end{tabular}%
    }
    \caption{Volume and Auction inflow rates during perimeter activation for different hourly flow limits. All values in veh/hr.}
    \label{tab:volume_auction_flow_limits}
\end{table*}

The volume-based approach is ineffective as it allocates green time based on the target inflow-to-capacity ratio, merely reducing the inflow’s green time share without ensuring a fixed vehicle flow. In contrast, the auction-based approach enforces a fixed vehicle count limit, providing more effective regulation.

There are potential improvements to be made with the auction-based approach. It mostly produced inflows close to the target intermittently disrupted by sharp increases, raising the mean. Elevated inflows in initial budget periods also contributed to the high maximum values in Table \ref{tab:volume_auction_flow_limits}. Further refinement is needed to keep inflows near target limits, potentially by gradually adjusting the flow limit instead of the abrupt step-change used in the case study.

\subsubsection{Delays} 
\begin{figure}[t]
    \centering
    \includegraphics[width=.35\textwidth]{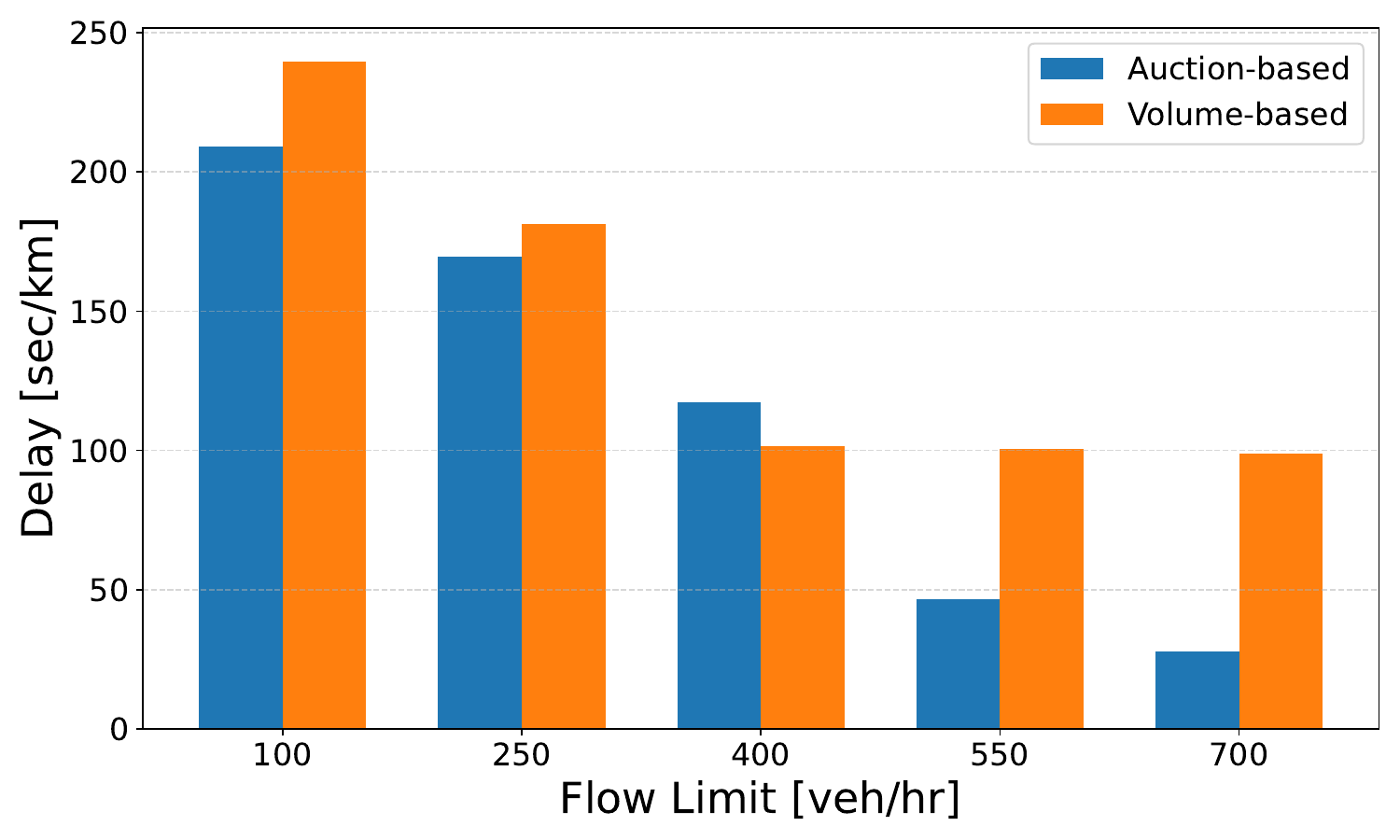} 
    \label{fig:Delays}
    \caption{Mean delays for vehicles for different flow limits.}
    \label{fig:MeanDelay}
\end{figure}

Figure \ref{fig:MeanDelay} shows that the auction-based approach generally results in lower mean vehicle delays than the volume-based approach. At 100–250 veh/hr, auction-based delays are lower when both methods influence gating. From 400–700 veh/hr, volume-based gating is ineffective, producing delays similar to no gating. This explains its lower delays at 400 veh/hr, where auction-based gating still restricts flows. As flow limits rise, auction-based delays decrease, as expected with more vehicles passing through. At 700 veh/hr, auction-based inflows resemble no gating (Table \ref{tab:volume_auction_flow_limits}), reflected in the minimal delays. Comparison at 700 veh/hr highlights the superiority of auction-based control over fixed-time control in intersection delays when gating is essentially not in effect, aligning with previous studies (e.g., \cite{Iliopoulou2022}). Future work could look into the distribution of these delays amongst the gated and non-gated movements.

\section{Conclusion}

This paper introduced a novel auction-based mechanism for green time allocation at signalized intersections, integrating dynamic user bids with flow limits to enhance perimeter control. The proposed SBSP auction framework outperformed volume-based methods in regulating PN inflows, demonstrating superior queue management, flow accuracy, and delay reduction across multiple flow limits.

Future improvements could allocate budgets to individual movements rather than entire inflows, enabling finer control over PN entry and spillbacks. Addressing the challenge of separating combined movements would allow selective deactivation for more precise regulation. Additionally, dynamically adjusting green time parameters based on real-time traffic conditions may further enhance flow accuracy.

\section*{Acknowledgements}
The authors confirm contribution to the paper as follows: study conception, analysis of results, manuscript preparation: A. Roocroft \& M. Rinaldi; model formulation and implementation: A. Roocroft. All authors have reviewed the results and approved the final version of the manuscript.


\end{document}